\documentclass[aps,twocolumn,nopacs,superscriptaddress,preprintnumbers,prb]{revtex4-2}
\usepackage{graphicx} % Required for inserting images
\usepackage{url}
\usepackage{color}
\usepackage[colorlinks=true,
            linkcolor = blue,
            urlcolor  = blue,
            citecolor = blue,
            anchorcolor = blue]{hyperref}
            
\usepackage{comment}
\usepackage{mathrsfs}
\usepackage{subfigure}
\footnotetext{These authors contributed equally to this work}

\begin{document}
\title{Phonon-mediated superconductivity in transition-metal trioxides XO$_3$ (X = Ru, Re, Os, Ir, Pt) under pressure}
\author{Aiqin Yang$^{\star}$}
\affiliation{MOE Key Laboratory for Non-equilibrium Synthesis and Modulation of Condensed Matter, Shaanxi Province Key Laboratory of Advanced Functional Materials and Mesoscopic Physics, School of Physics, Xi'an Jiaotong University, 710049, Xi'an, China}
\affiliation{School of Physical Science and Technology, and Collaborative Innovation Center of Suzhou Nano Science and Technology, Soochow University, 1 Shizi Street, Suzhou 215006, China}

\author{Xiangru Tao$^{\star}$}
\affiliation{MOE Key Laboratory for Non-equilibrium Synthesis and Modulation of Condensed Matter, Shaanxi Province Key Laboratory of Advanced Functional Materials and Mesoscopic Physics, School of Physics, Xi'an Jiaotong University, 710049, Xi'an, China}

\author{Yundi Quan}
\email{yundi.quan@gmail.com}
\affiliation{MOE Key Laboratory for Non-equilibrium Synthesis and Modulation of Condensed Matter, Shaanxi Province Key Laboratory of Advanced Functional Materials and Mesoscopic Physics, School of Physics, Xi'an Jiaotong University, 710049, Xi'an, China}

\author{Peng Zhang}
\email{zpantz@mail.xjtu.edu.cn\\}
\affiliation{MOE Key Laboratory for Non-equilibrium Synthesis and Modulation of Condensed Matter, Shaanxi Province Key Laboratory of Advanced Functional Materials and Mesoscopic Physics, School of Physics, Xi'an Jiaotong University, 710049, Xi'an, China}

%\footnote{$^{\star}$ The two authors contributed equally to this work}

\date{\today}
\begin{abstract}
A recent experiment by Shan {\it et al} [arXiv:2304.09011] found that rhenium trioxide ReO$_3$, a simple metal at the ambient pressure, becomes superconducting with a transition temperature as high as 17 K at 30 GPa. In this paper, we analyze the electron-phonon origin of superconductivity in rhombohedral ReO$_3$ in detail. In addition, we also conduct a high-throughout screening of isostructural transition-metal trioxides XO$_3$ in searching for potential pressure-induced superconductors. Totally twenty-eight XO$_3$ compounds have been studied, in which four candidates RuO$_3$, OsO$_3$, IrO$_3$ and PtO$_3$ are predicted superconducting with the transition temperatures of 26.4, 30.3, 0.9 and 2.8 K at 30 GPa, respectively. Both IrO$_3$ and PtO$_3$ stay superconducting even at the ambient pressure. In ReO$_3$, RuO$_3, $OsO$_3$ and IrO$_3$, the conduction electrons around the Fermi level are dominantly from the X-d and the O-2p orbitals, and their electron-phonon coupling originates from the lattice dynamics of both the heavier transition-metal-atom and the oxygen-atom. Inclusion of spin-orbital coupling would mildly suppress the transition temperatures of these transition-metal trioxide superconductors except RuO$_3$.
\end{abstract}

\maketitle

%\thefootnote

\section{Introduction}
Searches for materials of higher superconducting transition temperatures, $T_c$, constitute one of the most challenging problems in theoretical and experimental physics. However, the first discoveries of almost all-types of superconductors come from unexpected breakthrough in experiments, including the Bardeen-Cooper-Schrieffer (BCS) superconductors \cite{onnes1911}, the cuprates \cite{Bednorz.ZPB1986}, the iron based superconductors \cite{Kamihara.JACS2008}, the heavy fermion superconductors \cite{Steglich.PRL2021} and the twisted bilayer graphene \cite{Cao.Nature2018}.
Efforts in elevating $T_c$ progressed slowly for the BCS-type superconductors until the discovery of superconducting hydrides.
%In 1911, Onnes discovered $T_c$ = 4.2 K superconductivity in mercury, being the first superconductor in history. After 90 years in 2001, MgB$_2$ was found superconducting with $T_c$ = 39 K, being the highest superconducting transition temperature among all BCS-superconductors at the ambient pressure at present. 
The BCS-type superconducting hydrides of record high $T_c$ surged in the past decades \cite{Wang.PNAS2012, MaLiang.PRL2022, Duan.SciRept2014, Drozdov.Nature2015, Peng.PRL2017, Liu.PNAS2017, Geballe.Angew2018, Somayazulu.PRL2019, Drozdov.Nature2019, Yang.PBCM2024} due to two entwined reasons \cite{2021Roadmap}. First, the elaborated Migdal-Eliashberg approach together with the first-principle density functional theory (DFT) enable the highly accurate calculations of $T_c$ in materials. Second, the development in high pressure techniques provides possibilities to synthesize unprecedented compounds \cite{Pickett.RevModPhys2023}. CaH$_6$ was predicted superconducting with $T_c$ $\approx$ 220-235 K at 150 GPa \cite{Wang.PNAS2012}, then was confirmed with $T_c$ $\approx$ 215 K at 172 GPa by later experiment \cite{MaLiang.PRL2022}. H$_3$S was predicted superconducting with T$_c$ $\approx$ 204 K at 200 GPa \cite{Duan.SciRept2014}, which was also confirmed by later experiment \cite{Drozdov.Nature2015}. Rare earth hydrides were also proposed superconducting \cite{Peng.PRL2017}, for example LaH$_{10}$ with T$_c$ $\approx$ 274-286 K at 210 GPa, and then experimentally confirmed \cite{Liu.PNAS2017, Geballe.Angew2018, Somayazulu.PRL2019, Drozdov.Nature2019}. Recently, Dasenbrock-Gammon {\it et al.} \cite{Dias.Nature2023} claimed the astonishing discovery of superconductivity in lutetium-nitrogen-hydrogen with $T_{\text{c}}$ = 294 K at 1 GPa. Unfortunately most of the subsequent experimental and theoretical works cannot reproduce their results, \cite{Shan.CPL2023, Ming.Nature2023, Xing.NC2023, Cai.MRE2023, Salke.arxiv2023, DPeng.MRE2023, Tao.SciBull2023} therefore leaves the quest for ambient superconductivity remaining unsolved.

\begin{figure}[!ht]
    \centering
    \includegraphics[]{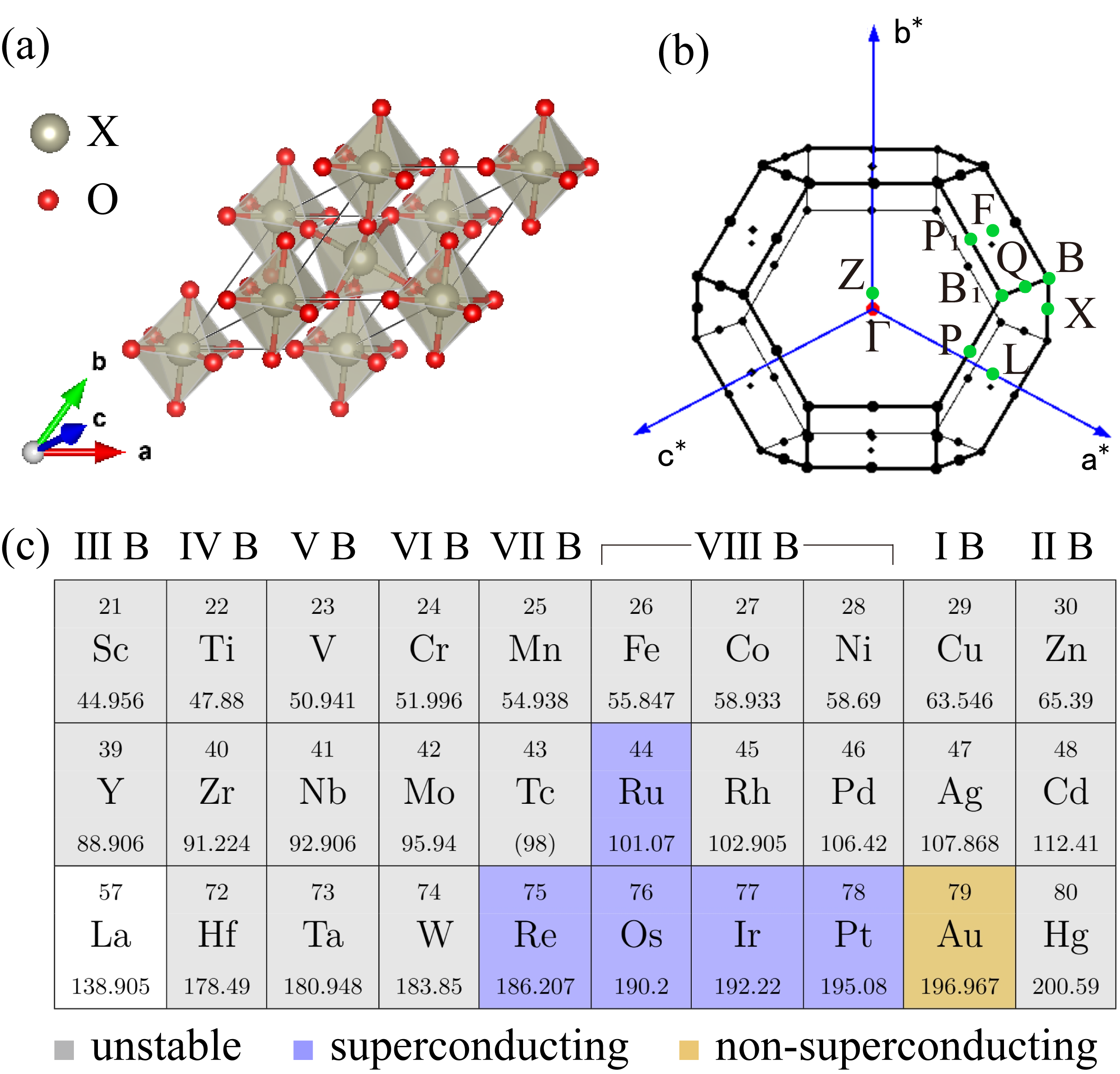}
    \caption{(a) The crystal structure of XO$_3$ ($R\bar{3}c$). (b) The first Brillouin zone of XO$_3$ ($R\bar{3}c$). (c) The transition-metal element X in the periodic table. The derived trioxides XO$_3$ ($R\bar{3}c$) at 30 GPa are  dynamically unstable in grey color, superconducting in purple color, and dynamically stable but non-superconducting in yellow color.}
    \label{fig:XO3-struct}
\end{figure}

Despite the high superconducting transition temperature of these hydrides, very high pressures are required not only to synthesize but also to stabilize these compounds. As stated above, the typical pressures are usually a few hundreds of GPa, which greatly limits the potential application of the superconducting hydrides. Searching for or even design new superconductors that may retain their superconductivity at relatively lower pressures are very attractive strategy. Recent experiment indicates ReO$_3$ of rhombohedral R$\bar{3}c$ symmetry (see Fig.~\ref{fig:XO3-struct}a) would have a maximum T$_c$ of 17 K at about 30 GPa \cite{Shan.arxiv2023}. To our knowledge, this is the highest T$_c$ among all known superconducting binary transition-metal oxides including YO \cite{YangQiuping.PhysRevMaterials.5.044802}, V$_2$O \cite{DuXin.PCCP2020}, NbO \cite{Hulm.JLTP1972}, TaO$_3$ \cite{LiWenjing.PCCP2023} and Ti$_n$O$_{2n-1}$ \cite{PhysRevB.104.224101}. Therefore, the experimental discovery of superconducting ReO$_3$ (R$\bar{3}c$) provides an interesting and rare example of superconducting transition-metal oxides at moderate pressure. Replacement of the Re-atom in ReO$_3$ (R$\bar{3}c$) by other transition-metal-atom may end up with potential transition-metal trioxide superconductors. Similar strategy of the metal-atom replacement in MgB$_2$-structures has lead to series of superconducting metal-borides \cite{Yu.PRB2022}, for example CaB$_{2}$ ($T_c$ $\sim$ 50 K \cite{Choi.PhysRevB.80.064503} or 9.4 - 28.6 K \cite{Yu.PRB2022} at 0 GPa, by theory), NbB$_2$ ($T_c$ $\sim$ 9.2 K at 0 GPa, by experiment )\cite{Schirber.PRB1992, Yamamoto.PCS2002, Takeya.PCS2004} 
, ScB$_2$ ($T_c$ = 1.5 K at 0 GPa, by experiment) \cite{Samsonov.1980} and MoB$_2$ ($T_c$ = 32 K at 100 GPa, by experiment) \cite{Pei.NSR2023}. Recently, ternary hydrides Mg$_2$IrH$_6$ was predicted superconducting with $T_c$ up to 160 K at the ambient pressure \cite{Sanna.npjcm.10.44(2024), Dolui.PhysRevLett.132.166001(2024)}. Later high-throughout screening of isostructrual X$_2$MH$_6$ compounds (X = Li, Na, Mg, Al, K, Ca, Ga, Rb, Sr, In, and M = 3d, 4d, and 5d transition-metals) successfully identified four potential thermodynamically meta-stable superconductors, Mg$_2$RhH$_6$, Mg$_2$IrH$_6$, Al$_2$MnH$_6$ and  Li$_2$CuH$_6$, whose $T_c$ are all estimated being above 50 K at the ambient pressure. \cite{Zheng.MTP.42.101374(2024)}

In this paper, we have firstly done thorough analysis on the origin of superconductivity in ReO$_3$ (R$\bar{3}c$). Then adopting ReO$_3$ of R$\bar{3}c$ symmetry as a template, we generated twenty-eight isostructural XO$_3$ compounds through the replacement of the rhenium-atom by the transition-metal-atom. We discovered four potential superconductors including OsO$_3$, RuO$_3$, IrO$_3$ and PtO$_3$ with superconducting transition temperature T$_c$ up to 33 K at 30 GPa. IrO$_3$ and PtO$_3$ stay dynamically stable even at the ambient pressure. The electronic structure, the phonon modes, and the electron-phonon coupling (EPC) of these XO$_3$-type superconductors are systematically analysed as well.
  
\section{Methods}
DFT calculations are carried out using the planewave code Quantum Espresso.\cite{Giannozzi_2009,Giannozzi_2017} The exchange-correlation potential is approximated using the generalized gradient approximation as parameterized by Perdew, Burke, and Ernzerhof (GGA-PBE).\cite{PBE} We use the optimized norm-conversing pseudopotential proposed by Hamann.\cite{ONCV,VANSETTEN201839} The kinetic energy cutoff and the charge density cutoff of the plane wave basis are chosen to be 80 and 320 Ry, respectively. Self-consistent calculations are carried out using a $\Gamma$-centered mesh with $24\times24\times24$ k-points and a Gaussian smearing width of 0.02 Ry. Structural optimization is performed as well, with the convergence threshold of total energy at $1.0\times 10^{-7}$ Ry and that on forces at $1.0\times 10^{-5}$ Ry/Bohr. The phonon dispersion, the EPC, and the superconducting transition temperatures are calculated using the density functional perturbation theory (DFPT) with a $4\times4\times4$ mesh of q-points. The superconducting transition temperature $T_c$ was estimated by the Allen-Dynes modified McMillan equation \cite{PhysRevB.12.905}:
\begin{eqnarray}
T_c = \frac{\omega_{log}}{1.2} \exp \left ( -\frac{1.04(1+\lambda)}{\lambda - \mu^\ast\left (1+0.62 \lambda 
 \right ) }\right )
 \label{eqn:Allen-Dynes-McMillan}
\end{eqnarray}
in which $\lambda$ is the average EPC parameter, $\omega_{log}$ is the logarithmic average frequency, and the Coulomb pseudopotential \cite{Morel.PR1962} $\mu^{*}$ = 0.1-0.13. Since some of the investigated transition-metals like rhenium have large atomic numbers, spin-orbit coupling (SOC) could be important for the electronic structures or the superconductivity. We have conducted calculations both with and without SOC in this paper to examine the effects of SOC. The Fermi surface nesting function of ReO$_3$ is derived by 
$\eta(\vec{q}) = \frac{1}{\Omega_{BZ}}\int_{BZ}dk \sum_{n,m} \delta(\epsilon_{kn}-E_F) \delta(\epsilon_{k+qm}-E_F)$, where $\Omega_{BZ}$ is the volume of the Brillouin zone, $E_F$ is the Fermi energy, and n, m represent the index of electronic bands. More details are referenced to the Supplementary Materials \cite{SI} (see also references \cite{USPEX,VASP, PAW, tetrahedron} therein).

\section{Results and discussion}

\subsection{Electronic structure of ReO$_3$ under pressure}
\begin{figure}[!ht]
    \centering
    \includegraphics[width=0.5\textwidth]{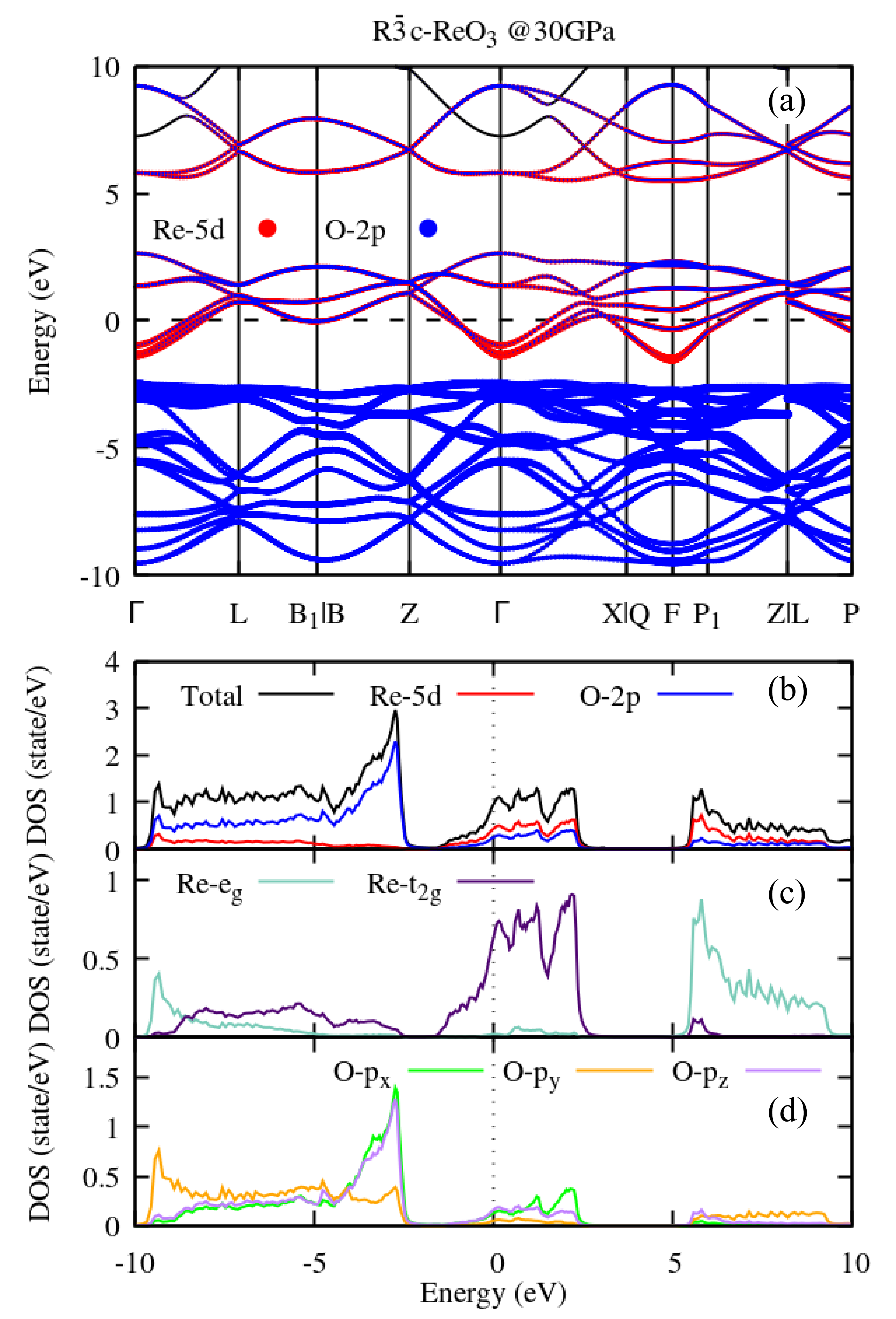}
    \caption{(a) The electronic band structure for ReO$_3$ ($R\bar{3}c$) at 30 GPa. (b-d) The total and the orbital projected density of states for Re-5d and O-2p at 30 GPa. SOC is not included.}
    \label{fig:ReO3-DOS}
\end{figure}

To elucidate the origin of superconductivity in the rhombohedral phase of ReO$_3$, we first presented the electronic band structure and the density of states (DOS) of ReO$_3$ ($R\bar{3}c$) in Fig.~\ref{fig:ReO3-DOS}. At pressure of 30 GPa, several energy bands cross the Fermi energy level indicating the metallic nature of ReO$_3$ ($R\bar{3}c$). The states below the Fermi level in range of [-10 eV, -2.5 eV] are dominantly occupied by the O-2$p$ electrons, together with minor contributions by the Re-5$d$ electrons. In contrast, the electronic DOS around the Fermi level originate from both the Re-5$d$ orbitals and the O-2$p$ orbitals. Due to the crystal fields in the ReO$_6$ octahedron, the Re-5$d$ orbitals split into the triply degenerated $t_{2g}$ orbitals as well as the doubly degenerated $e_g$ orbitals, while the O-2$p$ orbitals split into the $p_x$, $p_y$, and $p_z$ orbitals. The orbital projected DOS in local coordinate system as depicted in Fig.~\ref{fig:ReO3-DOS}(c-d) indicate the electrons near the Fermi energy level mainly occupy the $t_{2g}$ orbitals of rhenium and the $p_x$, $p_z$ orbitals of oxygen. Given the large atomic mass of rhenium-atom, the spin-orbit interaction in ReO$_3$ ($R\bar{3}c$) might be important. We also calculated the electronic structure of ReO$_3$ ($R\bar{3}c$) with SOC. As presented in Fig. S2, SOC only slightly changed the electronic structure of ReO$_3$ ($R\bar{3}c$) by partially lifting its band degeneracy.

\begin{figure}[!ht]
    \centering    
    \includegraphics[]{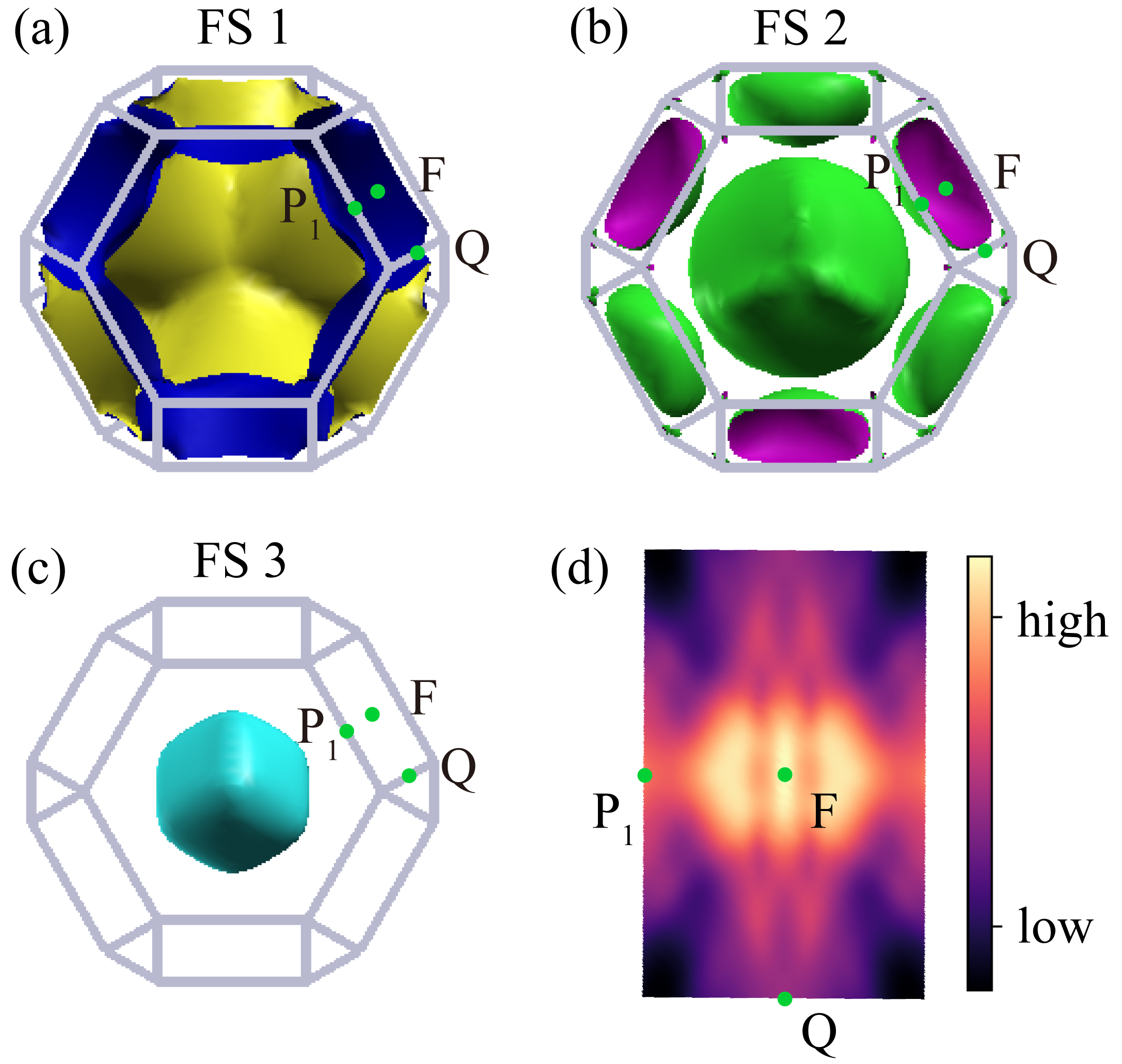}
    \caption{(a-c) The Fermi surfaces and (d) the Fermi surface nesting function $\eta(\vec{q})$ in the Q-F-P$_1$ plane of ReO$_3$ ($R\bar{3}c$) at 30 GPa. SOC is not included.}
    \label{fig:ReO3-FS-nesting}
\end{figure}

The Fermi surface of ReO$_3$ consists of three sheets labelled FS 1, FS 2 and FS 3 as shown in Fig.\ref{fig:ReO3-FS-nesting} (a-c). There are electron pockets around the $\Gamma$ point at the center of the Brillouin zone, electron pockets around the F point at the center of the tetragonal surface, as well as hole pockets around the Z and L points at the center of the hexagonal surface. The first Brillouin zone of ReO$_3$ with $R\bar{3}c$ symmetry is presented in Fig.\ref{fig:XO3-struct}b. Large sections of flat regions in the Fermi surfaces may lead to considerable contributions to the nesting function, and consequently the large EPC \cite{Yang.PRB2023}. Overall, the multiple flat regions in the Fermi surface of ReO$_3$ ($R\bar{3}c$) may lead to strong nesting. Therefore, we further calculated the Fermi surface nesting function of ReO$_3$ ($R\bar{3}c$) using the EPW code \cite{epw1,epw2} on a tetragonal plane defined by the Q-F-P$_1$ points as shown in Fig.\ref{fig:ReO3-FS-nesting}. There are nesting function peaks located at the F-point, as well as peaks on both sides of the F-point. These peaks of nesting function likely come from the electron pocket FS 2 around the $\Gamma$ point. 

\begin{figure*}[ht]
    \centering    \includegraphics[width=0.98\textwidth]{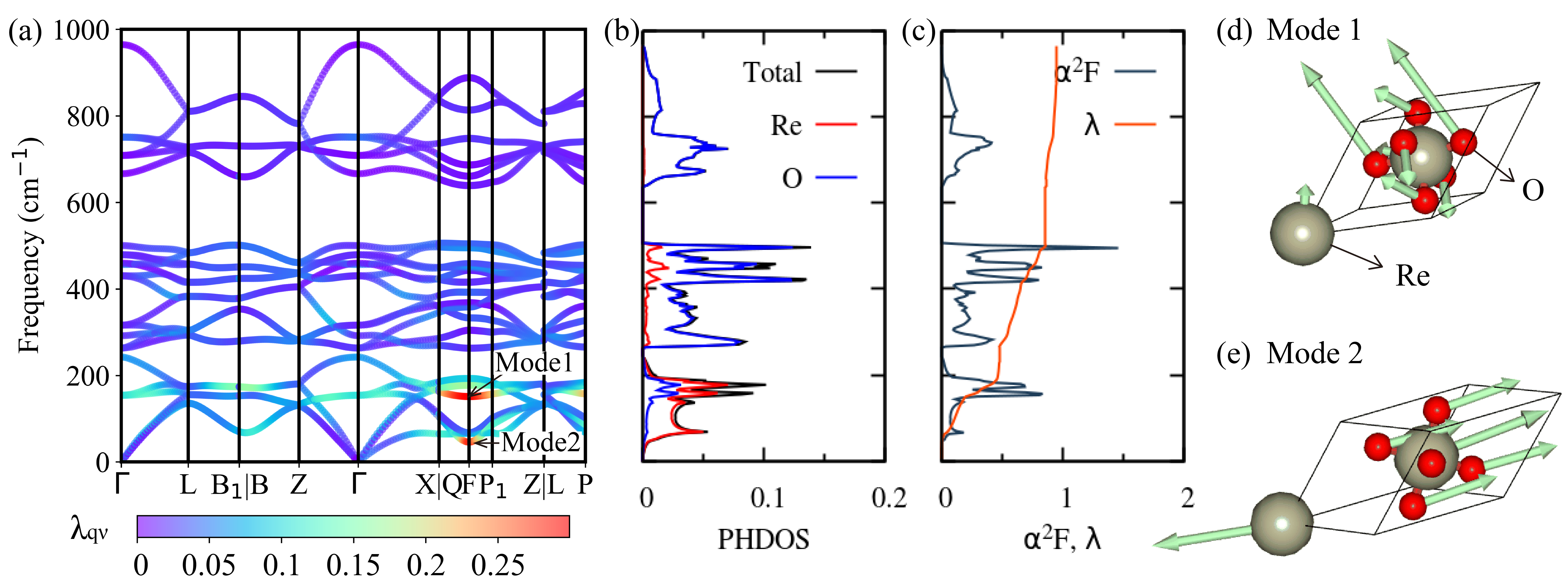}
    \caption{(a) The phonon dispersion weighted by the magnitude of EPC $\lambda_{q\nu}$. (b) The total and the atom-projected phonon DOS. (c) The Eliashberg spectral function $\alpha^2F$ and the cumulative frequency-dependent EPC function $\lambda$ of ReO$_3$ ($R\bar{3}c$) at 30 GPa. (d,e) The vibration modes corresponding to the two largest EPC $\lambda_{q\nu}$ in ReO$_3$ ($R\bar{3}c$). SOC is not included.}
    \label{fig:ReO3-elph}
\end{figure*}

\subsection{Superconductivity of ReO$_3$}
To investigate the EPC strength and the superconductivity of ReO$_3$ ($R\bar{3}c$), the phonon dispersion function, the phonon density of states (PHDOS), the Eliasheberg spectral function $\alpha^2F$, as well as the cumulative frequency-dependent EPC $\lambda(\omega)$ are calculated. The phonon dispersion function weighted by the magnitude of EPC $\lambda_{q\nu}$ (q is the wave vector and $\nu$ represents the index of phonon modes) is presented in Fig.~\ref{fig:ReO3-elph}a. There are no imaginary phonon modes in the dispersion function, clearly indicating the dynamically stability of ReO$_3$ ($R\bar{3}c$) at 30 GPa. In Fig.~\ref{fig:ReO3-elph}a, the phonon modes along the Q-F-P$_1$ path at around \textcolor{black}{150 cm$^{-1}$} have the most pronounced EPC $\lambda_{q\nu}$. This is consistent with our aforementioned analysis of the nesting function in Fig.~\ref{fig:ReO3-FS-nesting}d. The phonon mode at the F-point (mode 1) is associated with the vibrations of the rhenium- and the oxygen-atom as depicted in Fig.~\ref{fig:ReO3-elph}d. In mode 1, the six oxygen-atom at the octahedral vertex constitute three pairs that vibrate coherently in three different directions, while the vibrational magnitudes of the rhenium-atom are much smaller than these of the oxygen-atom. The next important phonon mode of large EPC $\lambda_{q\nu}$ locates at the F-point and around 46 cm$^{-1}$ (mode 2). In mode 2, four oxygen-atom at the octahedral vertex vibrate coherently in one direction, while the other two oxygen-atom stand still. The rhenium-atom at the center of the octahedron vibrates in direction close to, but not exactly parallel with, the vibration directions of the four oxygen-atom as discussed above.

The phonon spectrum of ReO$_3$ ($R\bar{3}c$) shows a wide range of frequency extending up to about 1000 cm$^{-1}$. From the PHDOS, it can be found that the vibrations of rhenium-atom dominate at the low frequencies, while the intermediate and high frequencies are mainly characterized by the vibrations of the oxygen-atom. The phonon modes at the low frequencies in range of 0-250 cm$^{-1}$ originate from the vibrations of both the rhenium-atom and the oxygen-atom, which count for about 49\% of the total EPC $\lambda$ = 0.95. The phonon modes at the intermediate frequencies in range of 250-500 cm$^{-1}$ mainly come from the vibrations of the oxygen-atom, accounting for about 40\% of the total EPC $\lambda$. The rest of the total EPC $\lambda$ originates from the phonon modes in the high-frequency region, being associated with the vibrations of the oxygen-atom.

The superconducting transition temperature $T_c$ is evaluated using the Allen-Dynes formula \cite{PhysRevB.12.905}. With the logarithmic average of the phonon frequencies $\omega_{log}$ = 347 K and the Coulomb pseudopotential $\mu^*$ = 0.13, we obtain $T_c$ of 18.9 K for ReO$_3$ ($R\bar{3}c$) as shown in Table ~\ref{Tab:SC}, which is close to the experimental value of 17 K \cite{Shan.arxiv2023}. When the SOC effect is taken into account, it can be found that  the PHDOS, the Eliashberg spectral function $\alpha^2F$, and the frequency-dependent EPC function $\lambda(\omega)$ almost don't change (see Fig. S3 in the Supplementary Materials \cite{SI}). The derived superconducting transition temperature including SOC decreases slightly to $T_c$ = 18.4 K in Table ~\ref{Tab:SC}, implying the minor effect of SOC on the superconductivity of ReO$_3$.  

\begin{table*}[ht]
\caption{The total electronic density of states (DOS) and the projected DOS of transition-metal-d and O-2p orbitals at the Fermi level $N(E_F)$ (unit: states/eV/f.u.), the average EPC parameter $\lambda$, the logarithmic average frequency $\omega_{log}$ and the superconducting transition temperature T$_c$ of RuO$_3$, OsO$_3$, IrO$_3$ and PtO$_3$ in $R\bar{3}c$ symmetry at 30 GPa without and with SOC.}
\resizebox{\textwidth}{!}{
\begin{tabular}{ccccccccccccc}
\hline
     & \multicolumn{6}{c}{without SOC} & \multicolumn{6}{c}{with SOC} \\ \cline{2-13} 
     & $N_{tot}(E_F)$ & $N_{X-d}(E_F)$ & $N_{O-p}(E_F)$ & $\lambda$   & $\omega_{log}$ (K)   & $T_c$ (K)         & $N_{tot}(E_F)$ & $N_{X-d}(E_F)$ & $N_{O-p}(E_F)$ & $\lambda$   & $\omega_{log}$ (K)   & $T_c$ (K)        \\ \hline
ReO$_3$ & 1.726 & 1.090 & 0.594 & 0.950    & 347     & 18.9-22.2      & 1.698 & 1.056 & 0.595 & 0.945   & 342    & 18.4-21.8     \\ 
RuO$_3$ & 2.708 & 1.340 & 1.257 & 2.542    & 168     & 26.4-27.9  & 2.685 & 1.328 & 1.251 & 2.224   & 204    & 29.5-31.4 \\ 
OsO$_3$ & 2.347 & 1.290 & 0.957 & 1.501    & 291     & 30.3-33.2      & 2.177 & 1.180 & 0.909 & 1.219   & 324    & 26.3-29.6 \\ 
IrO$_3$ & 1.641 & 0.717 & 0.864 & 0.401    & 478     & 0.9-1.4    & 1.563 & 0.675 & 0.835 & 0.373   & 484    & 0.5-1.3   \\ 
PtO$_3$ & 2.539 & 0.723 & 1.746 & 0.498    & 394     & 2.8-4.7    & 2.462 & 0.694 & 1.701 & 0.468   & 418    & 2.1-3.8   \\ 
\hline
\end{tabular}
}
\label{Tab:SC}
\end{table*}

\begin{figure*}[ht]    \includegraphics[width=\textwidth]{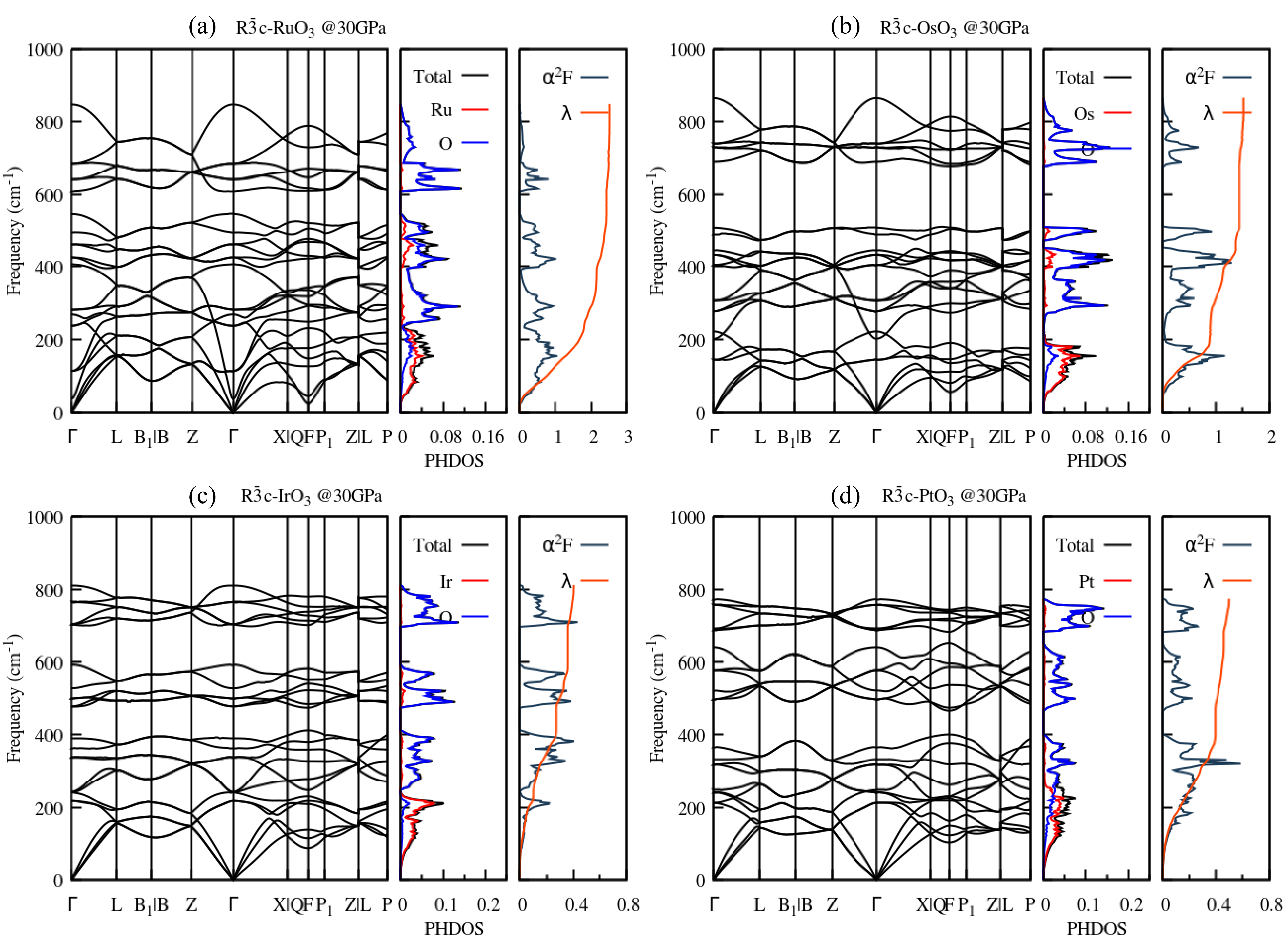}
    \caption{The phonon dispersion relation, the PHDOS, the Eliashberg spectral function $\alpha^2$F($\omega$) and the cumulative frequency-dependent EPC $\lambda (\omega)$ of RuO$_3$, OsO$_3$, IrO$_3$ and PtO$_3$ at 30 GPa. SOC is not included.}
    \label{fig:XO3-elph}
\end{figure*}

\subsection{Isostructrual transition-metal trioxides XO$_3$}

Inspired by the discovery of superconducting $T_c \sim$ 17 K in ReO$_3$ ($R\bar{3}c$) at 30 GPa \cite{Shan.arxiv2023}, we further investigated potential superconductivity in other isostructrual transition-metal trioxides. The electronic properties, the dynamical stability, and the superconducting transition temperatures of 28 XO$_3$-type transition-metal trioxides with $R\bar{3}c$ symmetry have been calculated by replacement of the rhenium-atom with the transition-metal-atom as listed in Fig.~\ref{fig:XO3-struct}c, where X represents all 3d, 4d, and 5d transition-metal-atom except lanthanides.

We calculated the phonon spectrum of all XO$_3$-type compounds to examine the dynamical stability of these systems. In addition to ReO$_3$, we identified five dynamically stable candidates with $R\bar{3}c$ symmetry, namely RuO$_3$ with 4$d$ electrons and OsO$_3$, IrO$_3$, PtO$_3$ and AuO$_3$ with 5$d$ electrons. The calculated phonon spectra of all five compounds at 30 GPa are shown in Fig.~\ref{fig:XO3-elph} as well as Fig. S14 in the Supplementary Materials \cite{SI}. The absence of negative frequencies indicates that these compounds are dynamically stable at 30 GPa. 

The derived $T_c$ of ReO$_3$, RuO$_3$, OsO$_3$, IrO$_3$ and PtO$_3$ by the Allen-Dynes equation are summarized in Table.~\ref{Tab:SC}. Four out of the five dynamically stable transition-metal trioxides, RuO$_3$, OsO$_3$, IrO$_3$ and PtO$_3$, are found superconducting at 30 GPa, while the derived superconducting transition temperature $T_c$ of AuO$_3$ is zero. Remarkably, the derived superconducting transition temperatures $T_c$ of RuO$_3$ and OsO$_3$ are 26.4-27.9 and 30.3-33.2 K at 30 GPa  respectively, which are even higher than that of ReO$_3$. In contrast, the derived superconducting transition temperatures $T_c$ of IrO$_3$ and PtO$_3$ are 0.9-1.4 K and 2.8-4.7 K at 30 GPa respectively, which are an order of magnitude lower than that of ReO$_3$. IrO$_3$ and PtO$_3$ would stay dynamically stable even at the ambient pressure as shown in Table S1 of the Supplementary Materials \cite{SI}. The $T_c$ of IrO$_3$ would increase to 8.4-11.4 K at the ambient pressure. In contrast, the $T_c$ of PtO$_3$ is 1.5-2.8 K at the ambient pressure, which is close to its $T_c$ at 30 GPa. Including of the spin-orbital coupling will slightly quench the $T_c$ except RuO$_3$ as presented in Table ~\ref{Tab:SC}. 

In Table ~\ref{Tab:SC}, RuO$_3$ is somewhat distinctive since it is compose of 4d transition metal ruthenium while the other four are composed of 5d transition metals. RuO$_3$ also has the largest EPC parameter $\lambda$ which may originate from its large electronic density of states at the Fermi level $N_{tot}(E_F)$ as well as the obvious phonon softening in its spectrum typically at the F point (see Fig.~\ref{fig:XO3-elph}a). However, the rather small logarithmic average frequency $\omega_{log}$ of RuO$_3$ encumbers the further elevation of its superconducting transition temperature T$_c$, leading to the second highest T$_c$ among the five superconducting transition-metal trioxides without SOC. Besides, inclusion of SOC would enhance its superconducting transition temperatures in contrast with the other four. 

To unveil the origin of superconductivity in these discovered transition-metal trioxide superconductors, we also calculated the phonon spectrum, the PHDOS, the EPC parameter $\lambda$ and the Eliashberg spectral function $\alpha^2F$ of RuO$_3$, OsO$_3$, IrO$_3$ and PtO$_3$. Our analysis suggest that the phonon modes of both the transition-metal-atom and the oxygen-atom have considerable contributions to the EPC and therefore the superconductivity of these four trioxides. As shown in Fig.~\ref{fig:XO3-elph}, the PHDOS of both the transition-metal-atom and the oxygen-atom in these four superconductors are mixed in the low and the intermediate frequency region.  In order to distinguish the respective contributions by the vibrations of the transition-metal-atom and the oxygen-atom to the total EPC parameter $\lambda$, we roughly separated the PHDOS into the low frequency region with dominant phonon modes from the transition-metal-atom and the high frequency region with dominant phonon modes from the oxygen-atom. For RuO$_3$, the PHDOS of Ru-atom and O-atom have the same magnitudes at 230 cm$^{-1}$. The $\lambda$ of RuO$_3$ accumulates quickly to about 1.918 at 230 cm$^{-1}$, which mainly comes from the EPC with the phonon modes of the Ru-atom and accounts for approximately 75\% of the total $\lambda$ = 2.542. Similarly, the PHDOS of Os-atom and O-atom in OsO$_3$ have the same magnitudes at 210 cm$^{-1}$. The $\lambda$ of OsO$_3$ also accumulates rapidly to 0.890 at 210 cm$^{-1}$, which mainly comes from the EPC with the phonon modes of the Os-atom and accounts for about 59\% of the total $\lambda$ = 1.501. In contrast, the $\lambda$ of IrO$_3$ and PtO$_3$ accumulate much slower relative to these of RuO$_3$ and OsO$_3$. The PHDOS of Ir- and Pt-atom dominant below approximately 260 cm$^{-1}$. The $\lambda$ of IrO$_3$ and PtO$_3$ increase to 0.105 and 0.225 at 260 cm$^{-1}$, which account for only 26\% and 45\% of their total $\lambda$ = 0.401 and 0.498, respectively. Clearly, the EPC being contributed by the transition-metal phonon dominant part in IrO$_3$ ($\lambda_{\text{Ir}}$ = 0.105) and PtO$_3$ ($\lambda_{\text{Pt}}$ = 0.225) are much smaller than these in RuO$_3$ ($\lambda_{\text{Ru}}$ = 1.918) and OsO$_3$ ($\lambda_{\text{Os}}$ = 0.890).  

IrO$_3$ and PtO$_3$ have small total EPC $\lambda$ at 0.401 and 0.498 respectively, which leads to their low superconducting transition temperature T$_c$. As shown in Table~\ref{Tab:SC}, $N_{X-d}(E_F)$ of IrO$_3$ and PtO$_3$ are around 0.7, being obviously smaller than these of ReO$_3$, RuO$_3$ and OsO$_3$. The low $N_{X-d}(E_F)$ in IrO$_3$ and PtO$_3$ suppressed the EPC among the transition-metal-d electrons and the phonon modes of Ir-, Pt-atom, which leads to the slow accumulation of $\lambda$ as being discussed in paragraph above. In comparison, the DOS of O-2p orbitals in IrO$_3$ and PtO$_3$ are comparable to these in ReO$_3$, RuO$_3$ and OsO$_3$. The EPC being contributed by the oxygen phonon dominant part are $\lambda_{\text{O}}$ = 0.296 and 0.273 in IrO$_3$ and PtO$_3$, which are relatively smaller than these of RuO$_3$ and OsO$_3$ at 0.624 and 0.611, respectively. Therefore, the small total $\lambda$ and the low T$_c$ in IrO$_3$ and PtO$_3$ mainly originate from the low electronic DOS of their transition-metal-d orbitals at the Fermi level.  \\

\begin{figure}[t]
    \centering
    \includegraphics[width=0.48\textwidth]{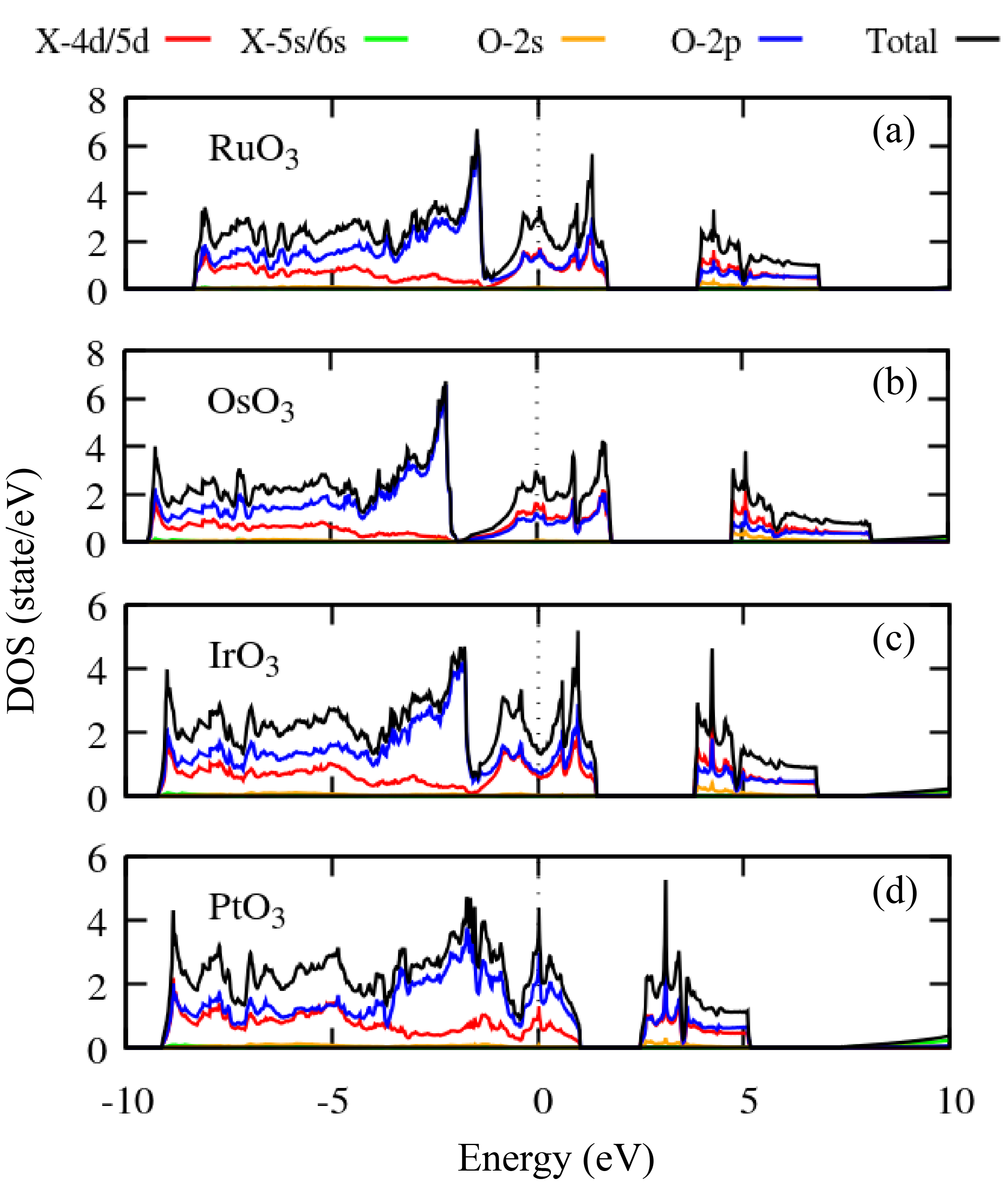}
    \caption{The total and the orbital projected electronic DOS of (a) RuO$_3$, (b) OsO$_3$, (c) IrO$_3$ and (d) PtO$_3$ at 30 GPa. SOC is not included.}
    \label{fig:XO3-dos}
\end{figure}

\subsection{Electronic structure and thermodynamically stability of XO$_3$ compounds}
The total and the projected electronic DOS of superconducting RuO$_3$, OsO$_3$, IrO$_3$ and PtO$_3$ are shown in Fig.~\ref{fig:XO3-dos}. In RuO$_3$, OsO$_3$ and IrO$_3$, the electronic DOS around the Fermi level are dominantly contributed by the Ru-4d, Os-5d, Ir-5d as well as the O-2p orbitals. In contrast, although in PtO$_3$ there are considerable electronic DOS from the Pt-5d orbitals around the Fermi level, the DOS from the O-2p orbitals are more than twice the magnitude. The electronic structures of these four trioxides indicate that the EPC with both the transition-metal-4d/5d electrons and the oxygen-2p electrons are important for their superconductivity, which is consistent with our analysis in previous sections. Similar scenario works for AuO$_3$ with near zero Au-5d electronic DOS around the Fermi level, which eliminates the possibility of superconductivity in AuO$_3$ (see Table S1 and Fig. S14 in the Supplementary Materials \cite{SI}). \\

We also have done thorough calculations of the formation enthalpy in the four transition-metal-oxygen binary systems, Ru-O, Os-O, Ir-O and Pt-O, as summarized in Fig. S1 of the Supplementary Materials \cite{SI}. Our calculations indicate that the four superconducting transition-metal trioxides are all thermodynamically metastable, in which RuO$_3$, OsO$_3$, IrO$_3$ and PtO$_3$ of R$\bar{3}c$ symmetry are 143, 62, 91 and 236 meV above the convex hull, respectively.

\section{Conclusion}
In summary, we have done thorough analysis of the electron-phonon origin of superconductivity in ReO$_3$ (R$\bar{3}c$). By high-throughout screening of the isostructural XO$_3$ (R$\bar{3}c$) compounds in which the rhenium-atom is replaced by 28 transition-metal-atom X, we found four potential superconductors RuO$_3$, OsO$_3$, IrO$_3$ and PtO$_3$ at 30 GPa with superconducting transition temperatures T$_c$ at 26.4-27.9, 30.3-33.2, 0.9-1.4  and 2.8-4.7 K, respectively. IrO$_3$ and PtO$_3$ stay dynamically stable even at the ambient pressure with T$_c$ equals to 8.4-11.4 and 1.5-2.8 K, respectively. In the five superconducting XO$_3$-type compounds, spin-orbit coupling only mildly suppressed their superconducting transition temperatures except RuO$_3$. In ReO$_3$, OsO$_3$, RuO$_3$ and IrO$_3$, the electronic DOS around the Fermi level consist of both the transition-metal-d states and the oxygen-2p states at roughly the same weight. In contrast, the major electronic DOS around the Fermi level in PtO$_3$ are from the oxygen-2p orbitals. The suppressed T$_c$ in IrO$_3$ and PtO$_3$ mainly originate from their low electronic DOS of the transition-metal-d orbitals at the Fermi level. The EPC and the superconductivity in all five XO$_3$ superconductors are contributed by the vibrations of both the heavier transition-metal-atom and the oxygen-atom. Our works thus shed lights on future experimental explorations for potential XO$_3$-type superconductors at relatively low or even the ambient pressures. \\

\section{Acknowledgement}
We would like to thank the support of the National Natural Science Foundation of China (Grants No.11604255), the Fundamental Research Funds for the Central Universities (Grants No.xzy022023011 and xhj032021014-04), and the Natural Science Basic Research Program of Shaanxi (Grants No.2021JM-001). The calculations were performed at the High Performance Computing Center of Xi’an Jiaotong University. \\
Aiqin Yang and Xiangru Tao contributed equally to this work.

\bibliography{refs.bib}
\end{document}